\documentclass[12pt,a4paper,final]{iopart}
\usepackage{iopams}  
\usepackage{graphicx}
\usepackage{epsfig}
\usepackage{figsize}
\usepackage[breaklinks=true,colorlinks=true,linkcolor=blue,urlcolor=blue,citecolor=blue]{hyperref}
\usepackage{cite}

\def\bra#1{\left\langle #1\right|}                   
\def\ket#1{\left| #1\right\rangle}                   

\begin{document}

\title[Parametric study for optimal performance of CCQDs]{Parametric Study for Optimal Performance of Coulomb-Coupled Quantum Dots}

\author{Kum Hyok Jong, Song Mi Ri, Chol Won Ri}

\address{Department of Physics, Kim Il Sung University, Ryongnam Dong, Taesong District, Pyongyang, DPR Korea}
\ead{kh.jong1025@ryongnamsan.edu.kp}

\date{\today}


\begin{abstract}
	
We study the optimal output power and efficiency of the three-terminal quantum heat engine with Coulomb-coupled quantum-dots (CCQD). It has been well known that in the weak coupling regime, two kinds of dominant transport mechanisms are sequential tunneling and cotunneling processes in CCQD. What process becomes dominant, which can be controlled by several parameters such as temperature difference, bias voltage, Coulomb interaction and tunneling parameters, is one of the key problems to determine the performance of the heat engine. We show the parametric dependence of the output power and coefficient and find the optimal performance of this CCQD heat engine through genetic algorithm.

\end{abstract}

\vspace{2pc}

\submitto{\JPCM}

\maketitle


\section{Introduction}

Recently quantum heat engine, which is regarded as a heat-to-work device using thermoelectric effects at the mesoscopic scale, attracts a great interest in both theoretically\cite{085428, 115414, 115415, NJP19} and experimentally\cite{N854, Comptes, N920, 3033}. Since Hicks and Dresselhaus\cite{Hicks1, Hicks2, Hicks3} showed that the thermoelectric figure of merit can be increased in low dimensions and Mahan and Sofo\cite{Mahan} showed that the best energy filters are also the best thermoelectrics, the research into this field has further become active. Quantum dot has both low dimension and discrete energy, and the energy level and coupling strength to the reservoir are controlled by means of the gate voltage as well\cite{QD}, thus the quantum dot has become an important candidate for developing high-efficient heat engines.

The earliest models of quantum-dot heat engines were two-terminal devices, which consisted of single quantum dot coupled to two reservoirs at different temperatures and different electrochemical potentials, and they're still studied\cite{N920, 032001, 245323, 155423, 3033, 206801, 052104, Report}. Recent experiment\cite{N920} demonstrated that this engine can achieve much higher efficiency than the classical heat engines. But in this two-terminal geometry, the heat and charge flow are carried by the same particles. For applications, materials need to have high electrical conductance while at the same time low thermal conductance. Thus the intimate coupling between the heat and charge flow in two-terminal devices poses a big problem\cite{N854}. 

To overcome this, multi-terminal heat engines have been proposed\cite{Comptes}. These engines separate the system into a conductor and environment and control the interaction between the two subsystems. The three-terminal heat engine with Coulomb-coupled double quantum-dots\cite{115414, 085428, N854, Comptes, NJP19, Report, CJP56, 125422} is one of the examples (\Fref{fig:model}). Especially peculiar drag effects such as Coulomb drag\cite{Coulomb_drag} and thermal drag\cite{thermal_drag} make this device curious.

Coulomb drag effect, early proposed in layered conductors\cite{Sov}, has been extended to various systems, especially coupled quantum wires\cite{Wire1, Wire2}, quantum dots,\cite{115414, 115415, 085428, N854, Comptes, Report, 076801, 196801, 045433, 052118} . So far, for high performance quantum heat engines, much attention has been dedicated to this effect. In three-terminal devices mentioned above, due to the Coulomb drag between the conductor and gate system, thermal fluctuations in the gate system induce charge current in the conductor, and the separation of heat and current flow may lead to higher efficiency. Hence the three-terminal heat engines have been investigated in great depth, theoretically and experimentally.

In Ref.\cite{085428}, they first suggested the three-terminal setup and showed the behavior of power and efficiency. According to them, the efficiency of this heat engine depends linearly on the bias voltage, thus it can, in principle, achieve the Carnot efficiency by increasing the bias. For simplicity, they assumed that the transport between the dot and reservoir is defined by sequential tunneling of single electrons. In the Coulomb blockade regime sequential tunneling is exponentially suppressed and the coherent cotunneling events dominate the transport\cite{245402, Matveev}. In Ref.\cite{115414}, they calculated the current and power of a three-terminal engine considering cotunneling rates by using a quantum master equation and compared the results with those from Keldysh Green's function formalism. 
In the case of weak coupling between the dot and reservoir, one can use the $T$-matrix-based master equation for the calculation of tunneling rates\cite{Bruus}. To include the coherent cotunneling events, the higher-orders in $H_T$ must be considered in the $T$-matrix and considering up to the second order in $H_T$ made this cotunneling effect reflected successfully \cite{Bruus, 115414, Koch}.

Also, many experimental researches have been carried out to realize these heat engines and peculiar effects. Recent experiment\cite{N854} presented the first experimental realization of such Coulomb-Coupled Quantum Dot (CCQD) heat engine. They confirmed the separation of charge and heat flow in this configuration, and showed that the direction of generated charge current can be manipulated by the gate voltages while leaving the direction of heat flow unchanged. In this experiment, they applied a small bias voltage to the conductor system and controlled the gate voltages of the two dots to measure the current. According to Ref.\cite{085428, Report}, the current and power depend not only on the gate and temperature difference, but also on the bias voltage. Thus appropriate tuning for the bias voltage and the other parameters may lead to higher power. 

Of course, there are many kinds of multi-terminal heat engines, besides two or three-terminal engines, but here we focus only on the three-terminal ones. Main attention will be provided to the boundary between the region where sequential tunneling is dominant and the region where cotunneling effect is dominant. It is well-known that there is no one-to-one correspondence between power and efficiency of three-terminal CCQD\cite{085428, 115414}, therefore, it is very important to find the optimal conditions in terms of power and efficiency. These conditions are determined by several parameters such as Coulomb interaction, bias voltage, tunneling parameters and temperature difference. Of course, the influence of temperature is somewhat trivial, that is, the higher temperature difference is, the higher power and efficiency are, as you can expect easily, and thus, we will pay attention to former three parameters. Well-known optimization methods are Genetic Algorithm (GA)\cite{optimal1}, Simulated Annealing\cite{optimal2}, Particle Swarm Optimization (PSO)\cite{optimal3}, etc., and now we will use GA for our purpose. This will provide us with Pareto front for high performance. 

This paper is organized as follows: In Sec. \ref{sec:model} we give a model hamiltonian and main formalisms for the calculation. And Sec. \ref{sec:discussion} and Sec. \ref{sec:conclusions} present our results and conclusive discussion. Finally, \ref{app:regularization} details some technical aspects of regularization for cotunneling rates.


\section{Model Hamiltonian} \label{sec:model}

\begin{figure}[h]
	   \centering
	   \SetFigLayout{2}{1}
	   \subfigure[]{\includegraphics[width=7cm]{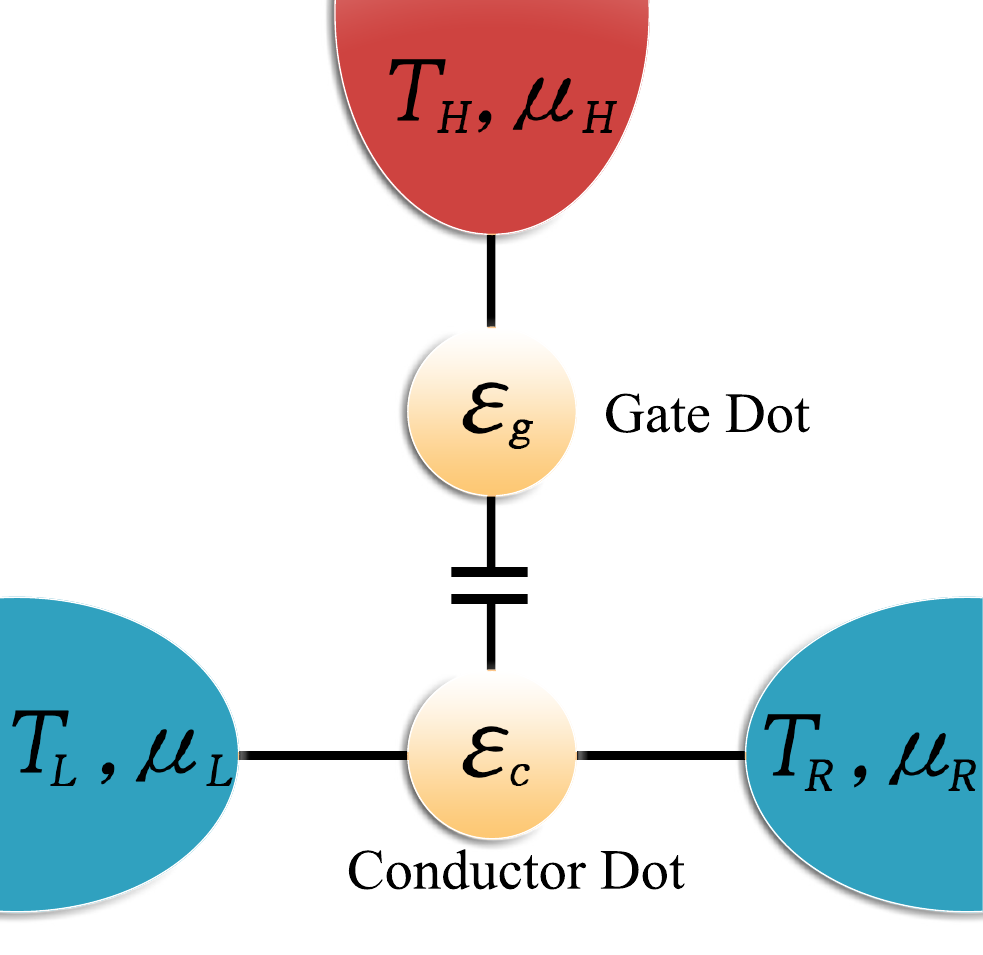}}
	   \subfigure[ ]{\includegraphics[width=10cm]{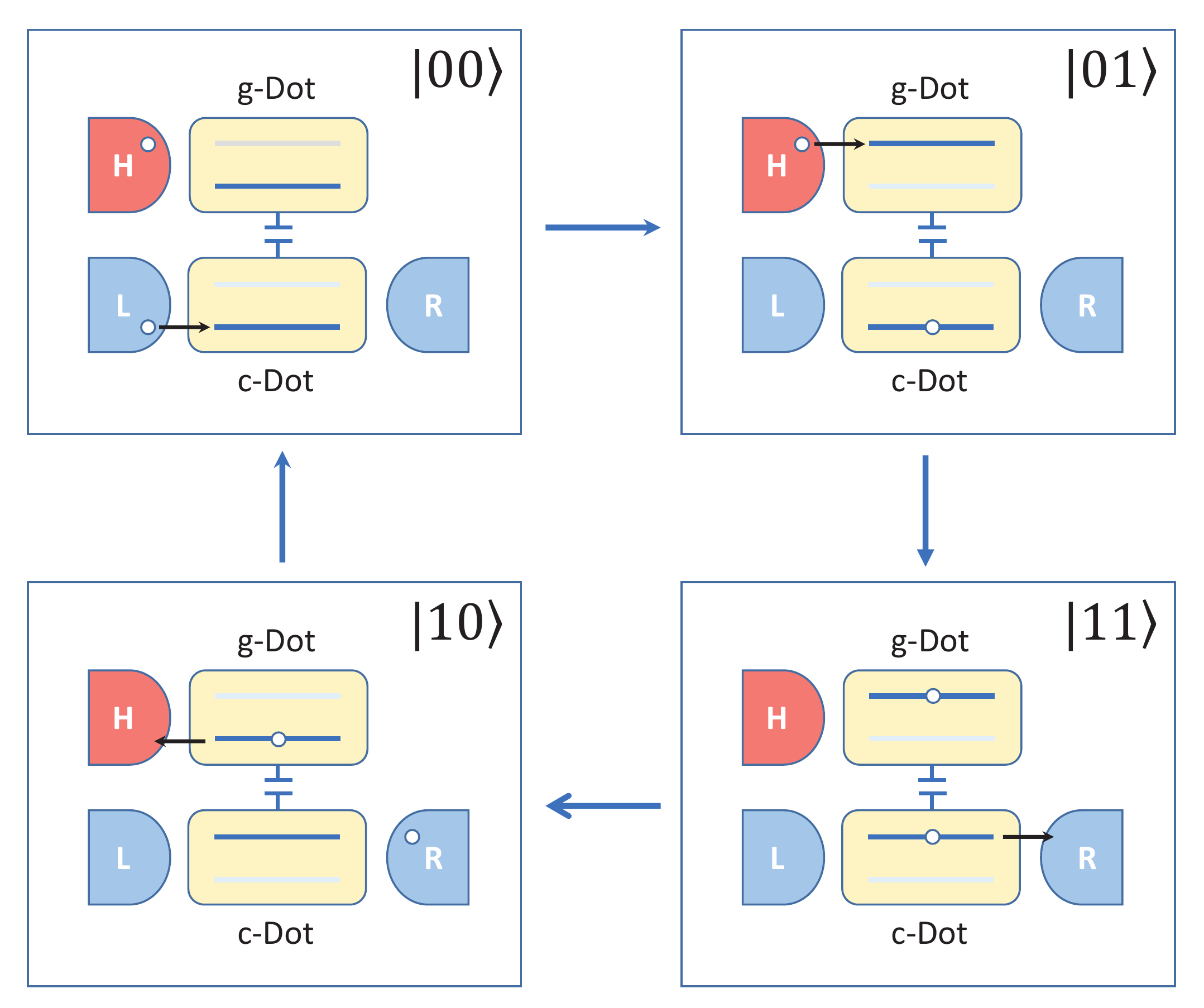}}
	   \caption{\label{fig:model}
			Schematic Picture of CCQD heat engine: (a) Model, (b) Operating principle. The energy $U$ is extracted from hot reservoir by inter-dot Coulomb interaction and generates the current from left to right reservoir against bias voltage.}
\end{figure}

The model of the three-terminal heat engine with CCQD is presented in \Fref{fig:model}, which is described theoretically and experimentally in many researches (e.g. see the Ref.\cite{085428, 115414, N854, Comptes}). Here, $g$ and $c$ denote the gate and conductor dot respectively, and the reservoirs coupled to them are indicated as $H$ (hot), $L$ (left) and $R$ (right). The temperatures and electrochemical potentials of the reservoirs are $T_\beta$ and $\mu_\beta (\beta=H,L,R)$. With the assumption of strong Coulomb interaction within the dot, the possibility that the dot has more than two extra electrons vanishes, thus we can ignore the spin of electrons\cite{085428}.

The Hamiltonian of this model is consisted of two parts: $H=H_0+H_T$, where the decoupled part of the dots and reservoirs $H_0$ reads

\begin{equation} \label{eq:Hamiltonian_0}
	H_0=\epsilon_g \hat{n}_g+\epsilon_c 
	\hat{n}_c+U\hat{n}_g\hat{n}_c+\sum_{\alpha=H,L,R}H_{0\alpha},
\end{equation}
where $H_{0\alpha}=\sum_k \epsilon_{k\alpha}\hat{n}_{k\alpha}$. Here, $\hat{n}_g=d^\dag_g d_g$, and $\hat{n}_c=d^\dag_c d_c$ describe the occupation numbers of the dot $g$ and $c$, with $d^\dag_g$ ($d_g$) and $d^\dag_c$ ($d_c$) indicating the creation (annihilation) operators of an electron with energy $\epsilon_g$ and $\epsilon_c$ in dot $g$ and $c$ respectively, and $U$ is the inter-dot Coulomb interaction. And $\hat{n}_{k\alpha}=c^\dag_{k\alpha} c_{k\alpha}$ is the occupation number of the state with energy $\epsilon_{k\alpha}$ and momentum $k$ in reservoir $\alpha$. The interacting part $H_T$ reads
\begin{equation} \label{eq:Hamiltonian_T}
	H_T=\sum_k (V_{kH}c^\dag_{kH}d_g+H.c.)+\sum_{\beta=L,R}\sum_k(V_{k\beta}c^\dag_{k\beta}d_c+H.c.),
\end{equation}
where $V_{k\alpha}=V_\alpha(\epsilon_{k\alpha})$ is the hybridization parameter\cite{5528} and $\Gamma_\alpha(\epsilon)=2\pi\rho_\alpha(\epsilon)\left|{V_\alpha(\epsilon)}\right|^2$ is the tunneling parameter, with the density of states of reservoir $\alpha$, $\rho_\alpha(\epsilon)$.

Fluctuations in the gate are maximized at half-filling\cite{115414}, thus we will consider the particle-hole symmetric case: $\epsilon_g=\epsilon_c \equiv \epsilon_d=-U/2$. Also, we choose the electrochemical potential of the top reservoir to $\mu_H=0$, and the bias voltage across the conductor system to be applied symmetrically ($\mu_L=-eV/2$, $\mu_R=eV/2$). The temperature of the top reservoir $T_H$ is higher than those of bottom reservoirs $T_L=T_R=T_C$. 

To break the left-right symmetry, we choose the tunneling parameters between dots and reservoirs as following step-like functions \cite{115414}.
\begin{eqnarray} \label{eq:Gamma}
\eqalign{
	\Gamma_L(\epsilon)=\frac{\Gamma_C}{\exp(\epsilon\beta_0)+1}  \cr
	\Gamma_R(\epsilon)=\frac{\Gamma_C}{\exp(-\epsilon\beta_0)+1}=\Gamma_L(-\epsilon),  
}
\end{eqnarray}
where $\frac{1}{\beta_0}\equiv\gamma$ is the width of the step-like function, while $\Gamma_H(\epsilon)=\Gamma_H$. In the limit of $\gamma\rightarrow0$($\beta_0\rightarrow\infty$), $\Gamma_L(\epsilon)=\Gamma_C\theta(-\epsilon)$, $\Gamma_R(\epsilon)=\Gamma_C\theta(\epsilon)$.

In the master equation, the states of many-body quantum systems are defined from the eigenstates when isolated from the reservoirs\cite{Report}. Thus from different occupations of the two dots, our system can have four possible states: $\ket{0}$ (both empty), $\ket{g}$ (only dot $g$ occupied), $\ket{c}$ (only dot $c$ occupied), $\ket{2}$ (both occupied). With the assumption of weak coupling between the dot and reservoir, the transition rate from state $\ket{m}$ to state $\ket{n}$ can be obtained from the generalized Fermi golden rule\cite{Bruus}.
\begin{equation} \label{eq:rate}
	\gamma_{mn}=\frac{2\pi}{\hbar}\sum_{i'f'}W_{i'}\left|\bra{f'n}T\ket{i'm}\right|^2\delta(E_f-E_i),
\end{equation}
where $\ket{i'}$, $\ket{f'}$ indicate the initial and final states of the reservoirs, and $\ket{i'm}$, $\ket{f'n}$ indicates the states of the whole system (dots and reservoirs), with the corresponding energy $E_i$, $E_f$. $W_{i'}$, the thermal probability of state $\ket{i'}$, is determined by the temperature and electrochemical potential of each reservoir. Considering up to the second order in $H_T$, the $T$-matrix expands as follows:
\begin{equation} \label{eq:T-matrix}
	T=H_T+H_T\frac{1}{E_i-H_0+i\eta}H_T, 
\end{equation}
where the first term corresponds to the sequential tunneling contribution, while the second term to the cotunneling one.

Using the master equation, the probability to find the system in state $\ket{m}$ reads
\begin{equation} \label{eq:master equation}
	\frac{d P_m}{d t}=\sum_{n\neq m}(\gamma_{nm}P_n-\gamma_{mn}P_m ),
\end{equation}
where the first part is the rate that the system arrives at the state $\ket{m}$, while the second part is the rate that the system leaves the state $\ket{m}$. Here, we focus the steady-state master equation $\frac{d P_m}{d t}=0$. According to the normalization condition $\sum_m P_m=1$ and particle-hole symmetry $\epsilon_g=\epsilon_c\equiv\epsilon_d=-U/2$, the probability of the four states lead to.
\begin{eqnarray} \label{eq:probability}
\eqalign{
	P_0=P_2=\frac{\hbar(\gamma_{g0}+\gamma_{c0})}{2(\Gamma_H+\Gamma_C)}  \cr
	P_g=P_c=\frac{\hbar(\gamma_{0g}+\gamma_{0c})}{2(\Gamma_H+\Gamma_C)}.   }
\end{eqnarray}
The charge current from the left and right reservoir to the bottom dot reads\cite{115414}
\begin{eqnarray} \label{eq:I_LR}
\eqalign{
	I_{L(R)}=&e[(\gamma_{0c}^{L(R)}+\tilde\gamma_{02}^{L(R)})P_0+(\gamma_{g2}^{L(R)}+\tilde\gamma_{gc}^{L(R)})P_g \cr
	&-(\gamma_{c0}^{L(R)}+\tilde\gamma_{cg}^{L(R)})P_c-(\gamma_{2g}^{L(R)}+\tilde\gamma_{20}^{L(R)})P_2], }
\end{eqnarray}
where the notation tilde is used for the cotunneling rates to distinguish from sequential ones. In the steady-state, the charge current through the conductor dot $I\equiv I_L=-I_R$ reads
\begin{eqnarray} \label{eq:I}
\eqalign{
	I=&I^{seq}+I^{cot},  \cr
	&I^{seq}=\frac{e}{2\hbar}\frac{1}{(\Gamma_H+\Gamma_C)}
	[\Gamma_H\Gamma_L(\epsilon_d)[f_L(\epsilon_d)-f_H(\epsilon_d)] \cr	&-\Gamma_H\Gamma_R(\epsilon_d)[f_R(\epsilon_d)-f_H(\epsilon_d)]+2\Gamma_L(\epsilon_d)\Gamma_L(\epsilon_d)[f_L(\epsilon_d)-f_R(\epsilon_d)]], \cr
	&I^{cot}=\frac{e \Delta}{2},
}
\end{eqnarray}
where $I^{seq}$ and $I^{cot}$ represent the sequence tunneling current and cotunneling current respectively and $f_{\alpha}$ is the Fermi function of reservoir $\alpha$($\alpha=H,L,R$). 
Due to the divergent part, the cotunneling current $e\Delta/2$(where $\Delta=\tilde{\gamma_{02}^L}-\tilde{\gamma_{02}^R}$, see \ref{app:regularization}) needs to be regularized. Two regularization methods have been proposed by Turek and Matveev\cite{Matveev} and Koch\cite{Koch}. In systems with a temperature bias, only the former method has been used, but we apply more precise method, the latter one to our model. The detailed regularization procedure is shown in \ref{app:regularization}.

Since the electron from left reservoir to conductor dot has the energy of $\epsilon_d$, the heat flow from left reservoir to conductor dot is $J_L=\frac{\epsilon}{d}I_L$. Analogically, the heat flow from right reservoir to conductor dot is $J_R=\frac{\epsilon_d+U}{e}I_R$. In the steady state, the condition $J_H+J_L+J_R=0, I_L=-I_R$ is satisfied, so the heat flow from the hot reservoir to the conductor reservoirs reads\cite{115414}
\begin{equation} \label{heat current}
	J_H\equiv J=\frac{U}{e} I.
\end{equation}
Thus the efficiency, the ratio of generated power $P=I V$ to the absorbed heat flow $J$, reads
\begin{equation} \label{efficiency}
	\eta=\frac{P}{J}=\frac{eV}{U}.
\end{equation}


\section{Result and Discussion}\label{sec:discussion}

Based on the many research results of three-terminal device, at first, we study the power and efficiency of our three-terminal device for several parameters to find the proper region for optimization process. All parameters become dimensionless with the temperature of hot reservoir.

\begin{figure}[t]
	\centering
	\SetFigLayout{1}{2}
	\subfigure[ ]{\includegraphics[width=7cm]{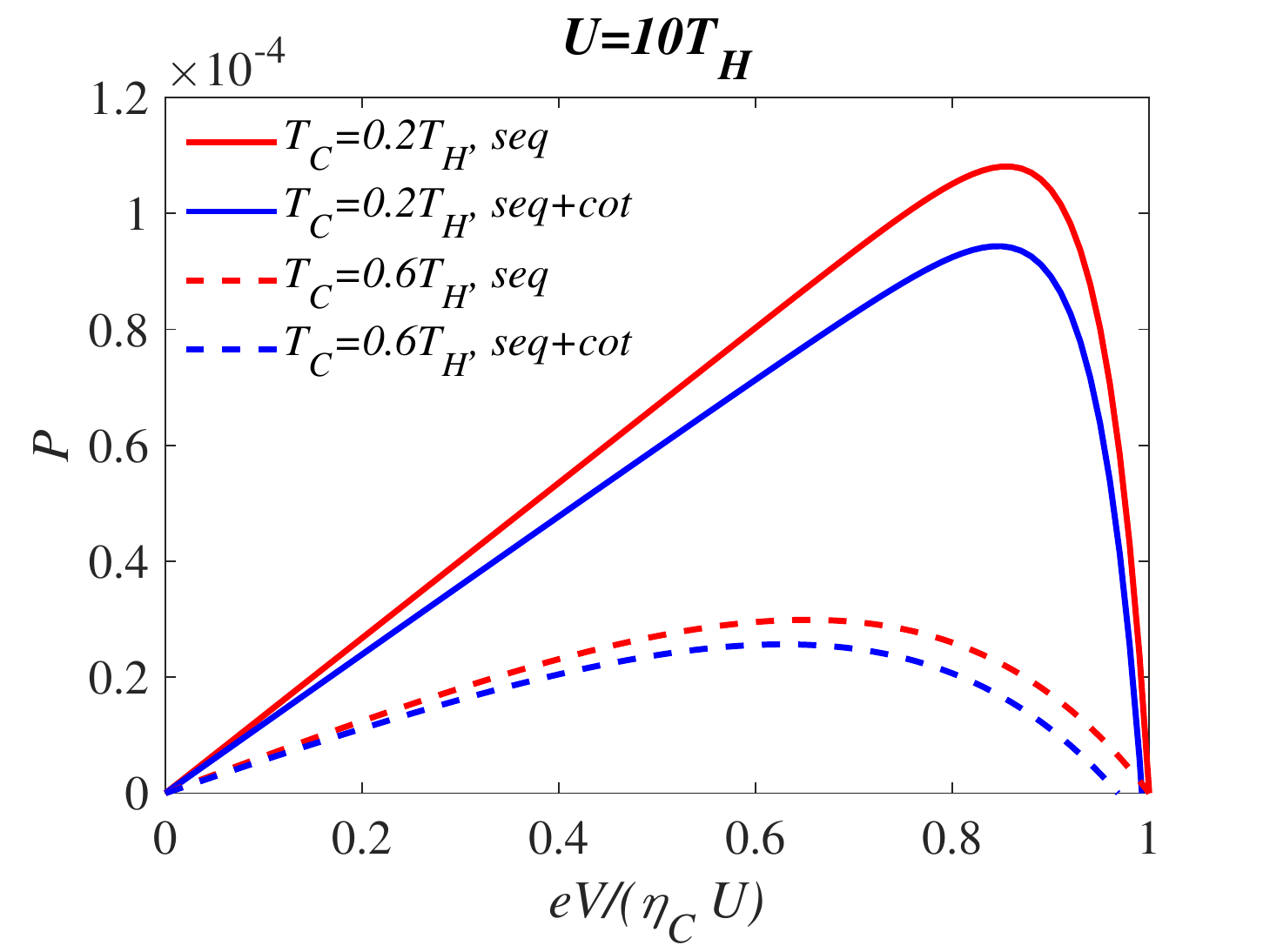}}
	\subfigure[ ]{\includegraphics[width=7cm]{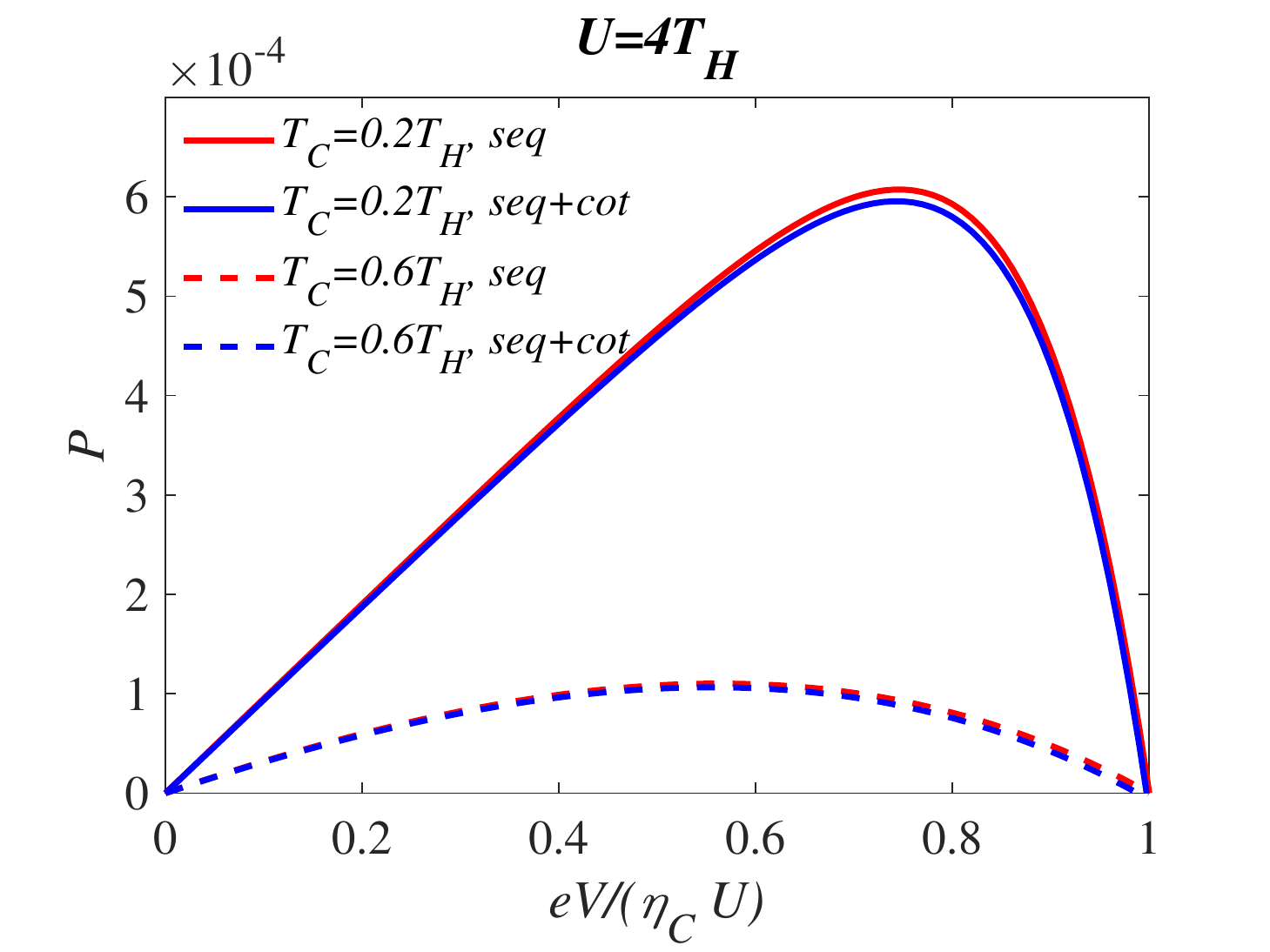}}
	\caption{\label{fig:P_V}
		Generated power as a function of the bias voltage. Red lines indicate the sequential contribution alone, and the blue lines include the sequential and cotunneling contributions. Parameters: $\Gamma_H=\Gamma_C=0.01T_H$, (a) $U=10T_H$, (b) $U=4T_H$.  }
\end{figure}

\begin{figure}[t]
	\centering
	\SetFigLayout{1}{2}
	\subfigure[ ]{\includegraphics[width=7cm]{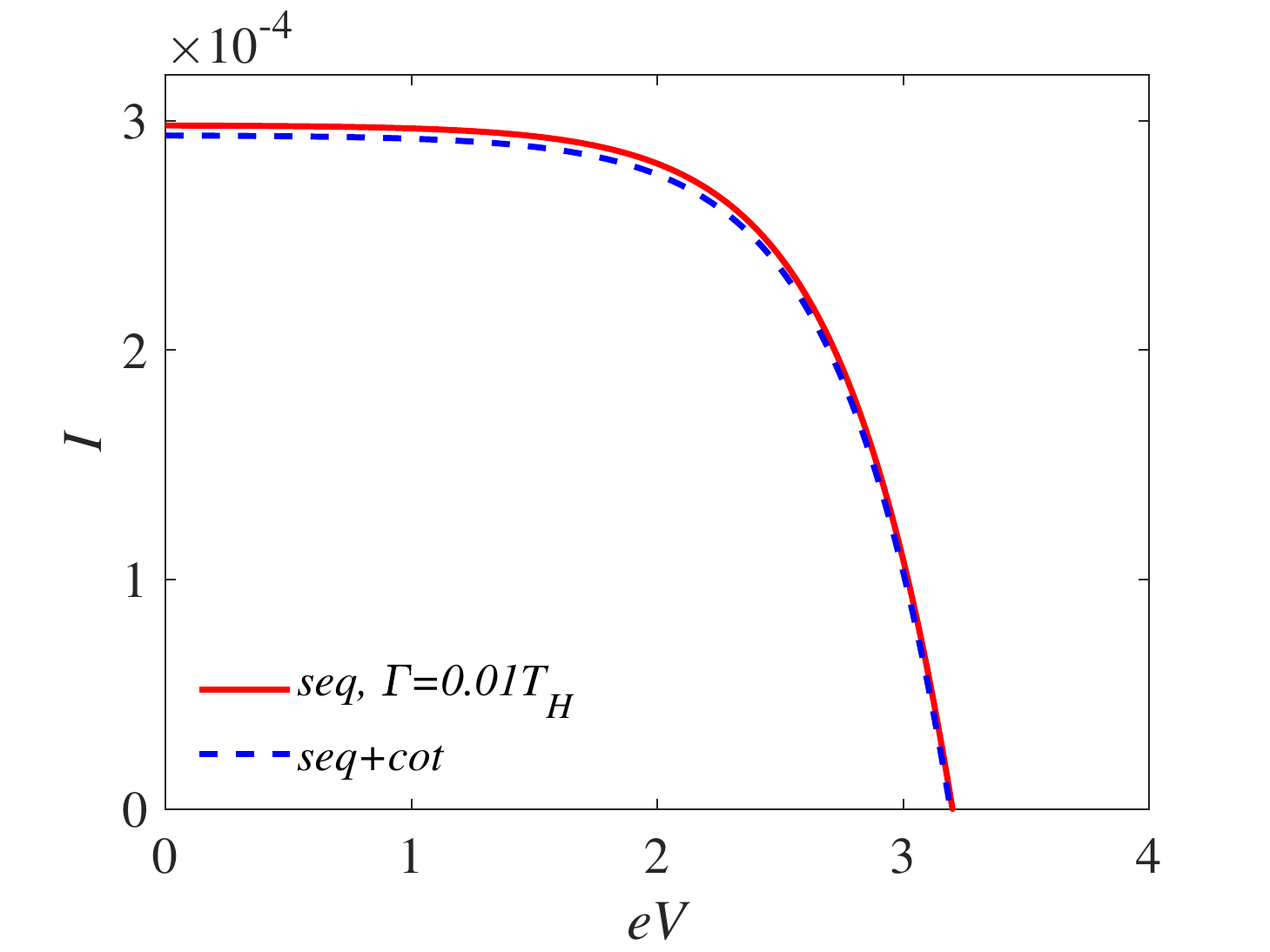}}
	\subfigure[ ]{\includegraphics[width=7cm]{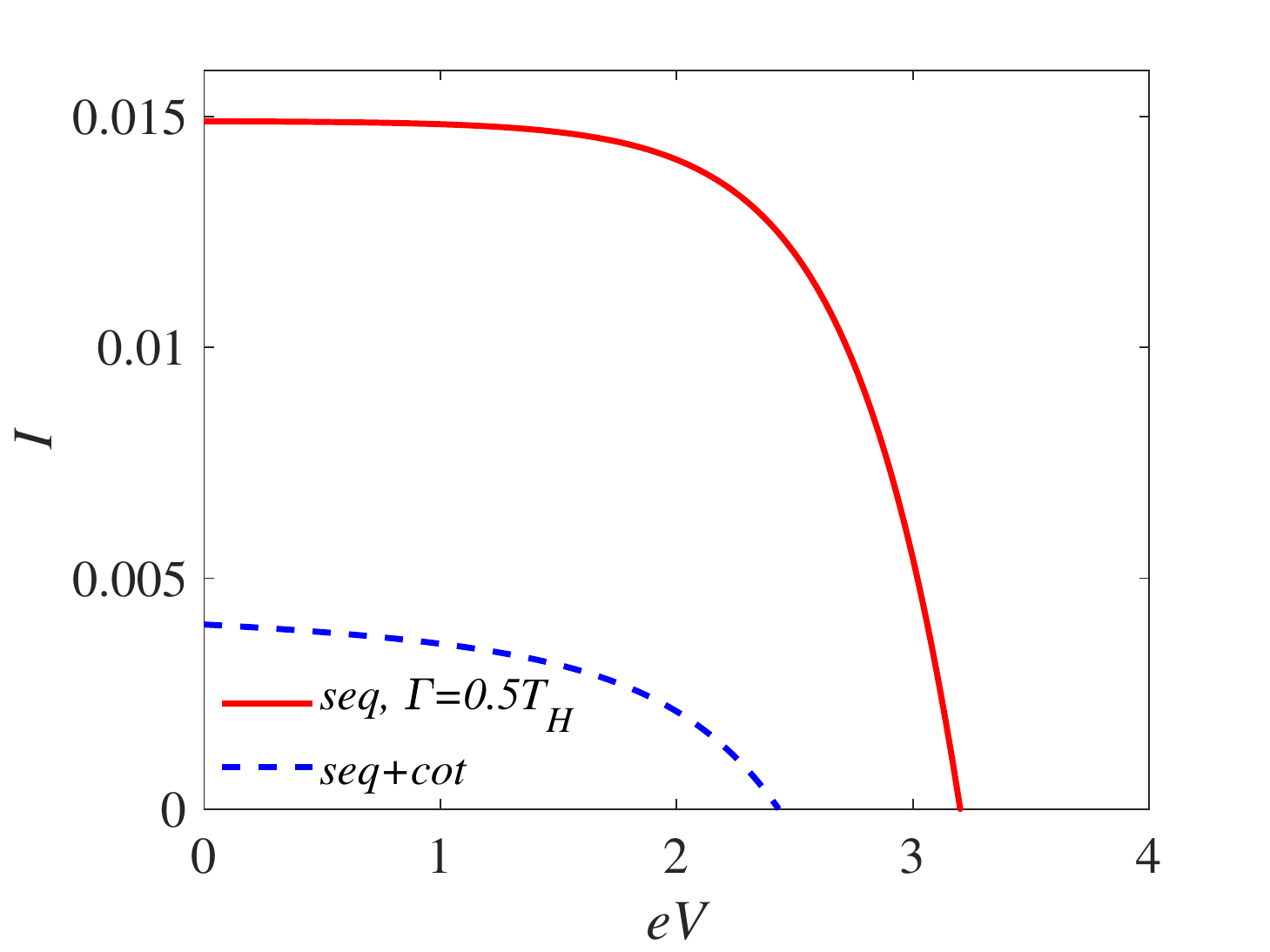}}
	\caption{\label{fig:I_V}
		Charge current through the conductor dot as a function of the bias voltage. Parameters: $T_C=0.2T_H$, $U=4T_H$, (a) $\Gamma=0.01T_H$, (b) $\Gamma=0.5T_H$.  }
\end{figure}

Besides small quantitative differences, our result coincides with Ref.\cite{085428, 115414, Report}, as shown in \Fref{fig:P_V} and \Fref{fig:I_V}. Increasing the bias voltage, the power first increases, but then decreases, defining the maximum power $P_{max}$ and stopping voltage $V_{stop}$. Comparing \Fref{fig:P_V}. (a) and (b), to begin with, we can find some interesting problems. In \Fref{fig:P_V}. (a), the contrast between the red line, indicating the sequential process, and blue line, containing cotunneling contribution, is remarkable due to the large Coulomb interaction: $U=10T_H$. For such a large $U$, this system lies in Coulomb blockade regime thus sequential tunneling processes decrease exponentially and cotunneling processes become dominant. In \Fref{fig:P_V}. (b), however, there is little difference between red and blue line. This indubitably indicates fine tuning of proper parameters will let the device to approach the optimal operating point. It is more important for the heat engines which have no one-to-one correspondence between power and efficiency. 

\Fref{fig:I_V} shows the I-V characteristics at $U=4T_H$ and different tunneling parameters, $\Gamma=0.01, 0.5T_H$, respectively. Below the Coulomb blockade regime, sequential tunneling process becomes dominant and makes the occupation number to change with the order of $\ket{00}\rightarrowtail\ket{01}\rightarrowtail\ket{11}\rightarrowtail\ket{10}\rightarrowtail\ket{00}$(see \Fref{fig:model}. (b)), while the first number represents the conductor dot's occupation number and second number is gate dot's one. So, it is possible to make the positive current against bias voltage. However, in the Coulomb blockade regime, the cotunneling process becomes dominant to fluctuate occupation number $\ket{01}\leftrightarrows\ket{10}$ or $\ket{00}\leftrightarrows\ket{11}$. This fluctuation makes the negative current according to the bias voltage, thus in this regime cotunneling process should be seriously considered. The cotunneling process depends on tunneling parameters as well as Coulomb interaction, as shown in \Fref{fig:I_V}. (a) and (b). As you can see, cotunneling effect becomes larger in the case of bigger tunneling parameter. This is because the tunneling parameter affects the fluctuation in the dot, alias, the stronger tunneling effect is, the larger fluctuation is. This is the reason why we select the tunneling parameter important for optimization process. Note that, this result differs with \cite{115414} from the difference of regularization method. However, this difference does not affect the result qualitatively, because it only exists in small bias voltage regime. As you will see later, this regime becomes out of our focus by selecting proper region for rapid optimization process. 

Of course, the temperature difference is also important, but its effect is somewhat trivial and we can find it by intuition through \Fref{fig:P_V}, i.e. the more temperature difference is, the more output power is and the higher efficiency is.  

\begin{figure}[t]
	\centering
	\SetFigLayout{2}{2}
	\subfigure[ ]{\includegraphics[width=7cm]{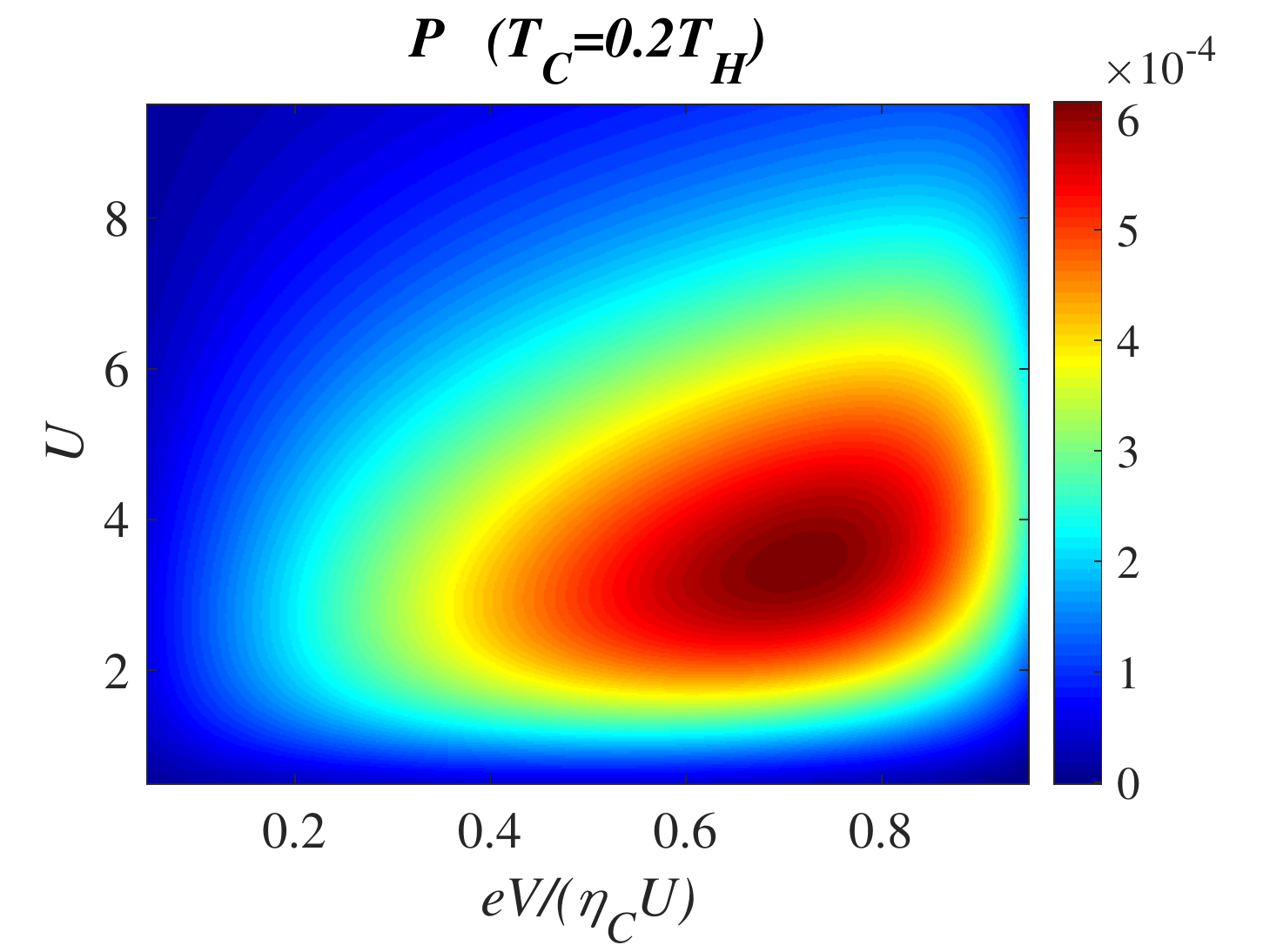}}
	\subfigure[ ]{\includegraphics[width=7cm]{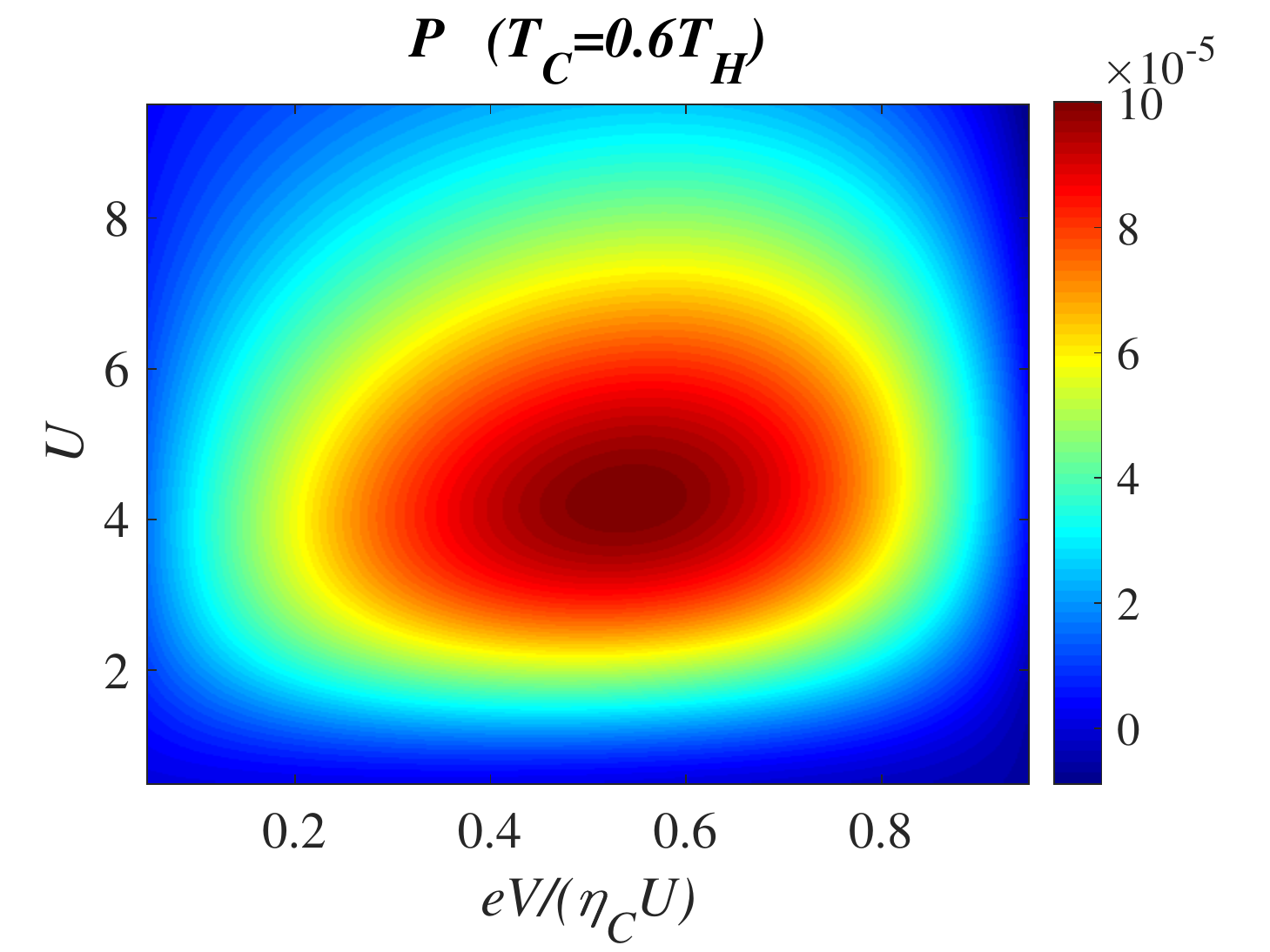}}
	\subfigure[ ]{\includegraphics[width=7cm]{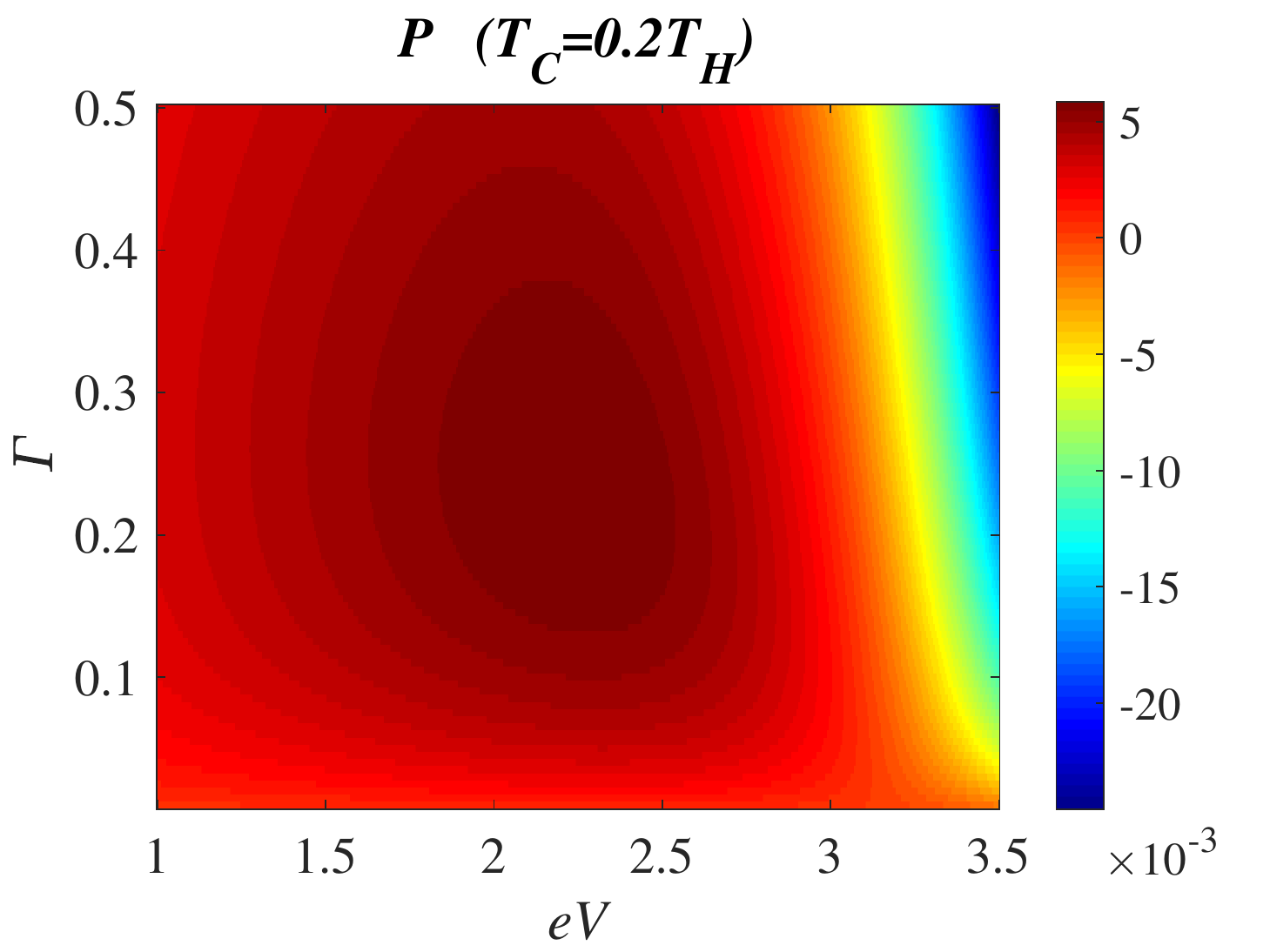}}
	\subfigure[ ]{\includegraphics[width=7cm]{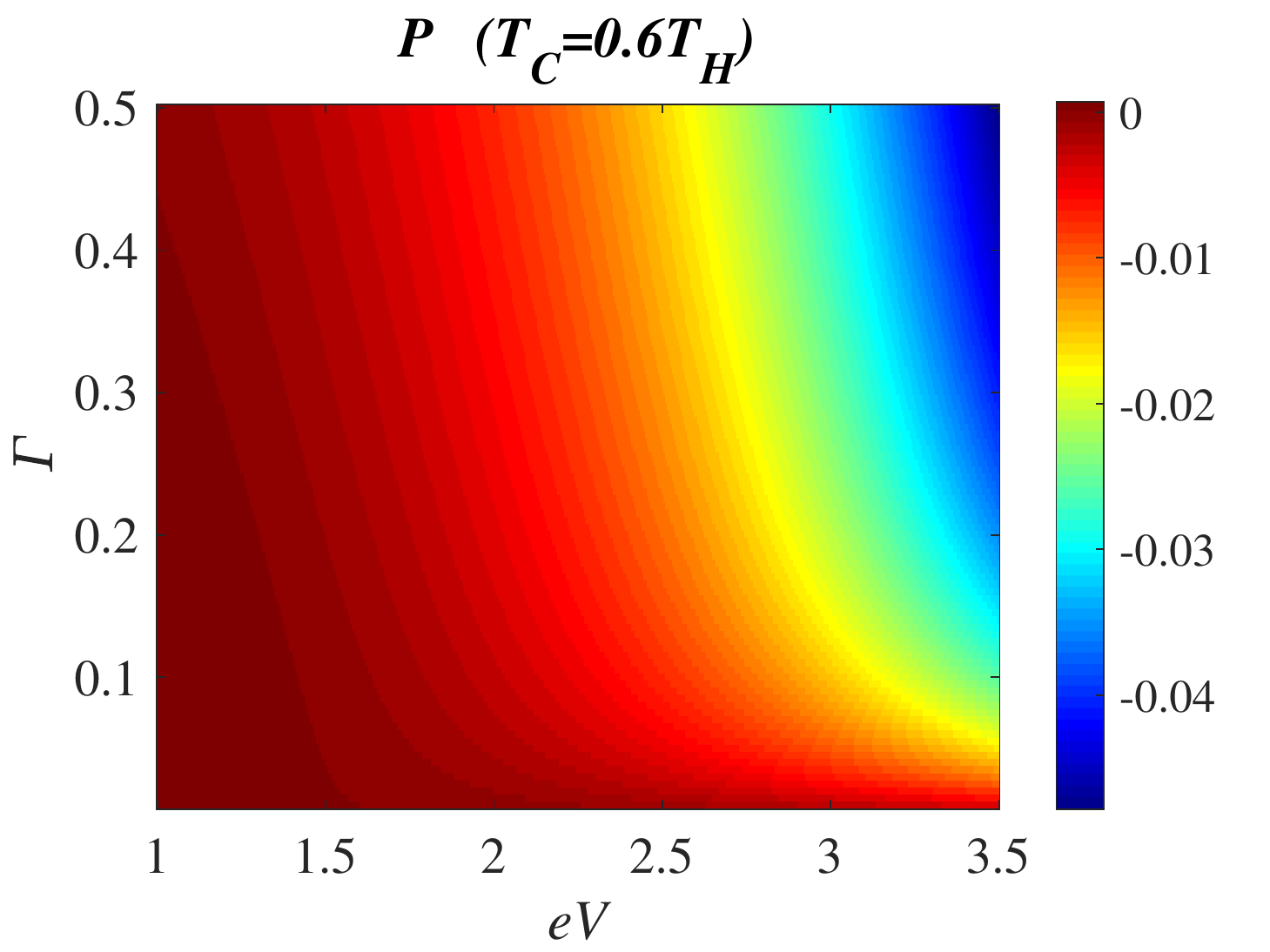}}
	\caption{\label{fig:P-V-U}
		Power as a function of the bias voltage and Coulomb interaction at different temperatures: (a) $T_C=0.2T_H, \Gamma=0.01T_H$, (b)$T_C=0.6T_H, \Gamma=0.01T_H$, and power as a function of the bias voltage and tunneling parameter at different temperatures: (c) $T_C=0.2T_H, U=4T_H$ (d) $T_C=0.6T_H, U=4T_H$}
\end{figure}

Now, we can set the multi-objective optimization problem for above CCQD heat engine. Generally, this problem can be modeled mathematically as following\cite{optimal1}:
\begin{eqnarray} \label{eq:moop}
\eqalign{
	max: &{f_1(\vec{x}),f_2(\vec{x}),...,f_n(\vec{x})}  \cr
	s.t: &g_i(\vec{x})\leq 0 \cr
	&\vec{x}_l\leq \vec{x} \leq \vec{x}_u \cr
}
\end{eqnarray}
where, $f_i(\vec{x})$ is the objective function, $g_i(\vec{x})$ is the constraint function and $\vec{x}$ is the vector consisted of concerning parameters, while $\vec{x}_l$ and $\vec{x}_u$ are the lower and upper limit of parameters, respectively. There are many powerful and intelligent algorithms for this multi-objective optimizing problem and we choose one of them, multi-objective genetic algorithm (MOGA). In this case, therefore, we choose three parameters as chromosome parameters in genetic algorithm, i.e. Coulomb interaction $U$, tunneling parameter $\Gamma$ and bias voltage $eV$, and the objective functions are output power and efficiency of CCQD heat engine. There is no constraint in this case. GA is evolutionary and heuristic search algorithm that is based on the basic principles of natural evolution\cite{optimal1}. Optimal solution has evolved from parents' population (a set of points in parameter space) to next generation  (known as children) with their own fitness through following steps; selection, crossover and mutation. Here, fitness is a function showed the differences between the parents and their expected optimal points. Selection is the process to select the two parent chromosomes from a population in terms of their fitness, which is assuming that the better parent will produce better offspring. Crossover is the process to form a new offspring with a certain crossover probability over the parents and mutation is the process to mutate new offspring at each chromosome with a certain mutation probability. This generation is repeated until the end condition is satisfied, and return the best solutions in current population. 

MOGA is a sort of GA for multi-objective problems and in generally, the objective functions in multi-objective problem always contradict each other. The improvement of one objective function may result decreasing of others, therefore, it is impossible to optimize all objective functions simultaneously. In this case, the corresponding optimal solutions are certain compromise between those objective functions, while they get close to their optimum as much as possible. These corresponding optimal solutions are called Pareto optimal solution and the set of these optimal solutions in parameter space is called Pareto front\cite{pareto287}. Theoretically, Pareto solutions are only acceptable in multi-objective problem. 
   
GAs are now rather widely used for solving optimization problems, however, the difficulty in their use is to find the objective function rapidly and exactly; moreover, same parents' population are selected repeatedly through the evolution process in the case of wide range of chromosome parameters. To avoid this difficulty, we should make the range of parameters as narrow as possible. 

\begin{figure}[t]
	\centering
	\SetFigLayout{1}{2}
	\subfigure[ ]{\includegraphics[width=7cm]{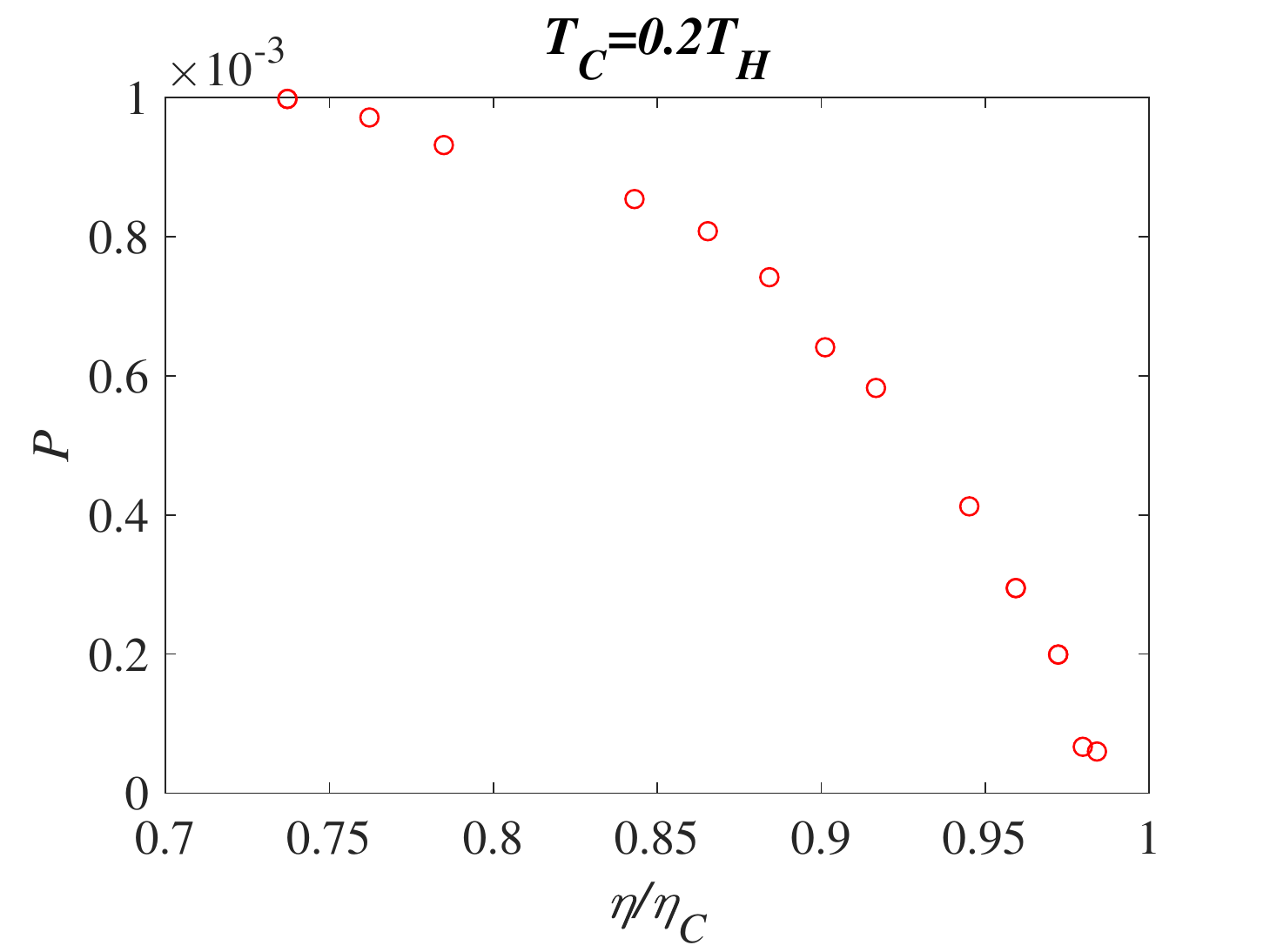}}
	\subfigure[ ]{\includegraphics[width=7cm]{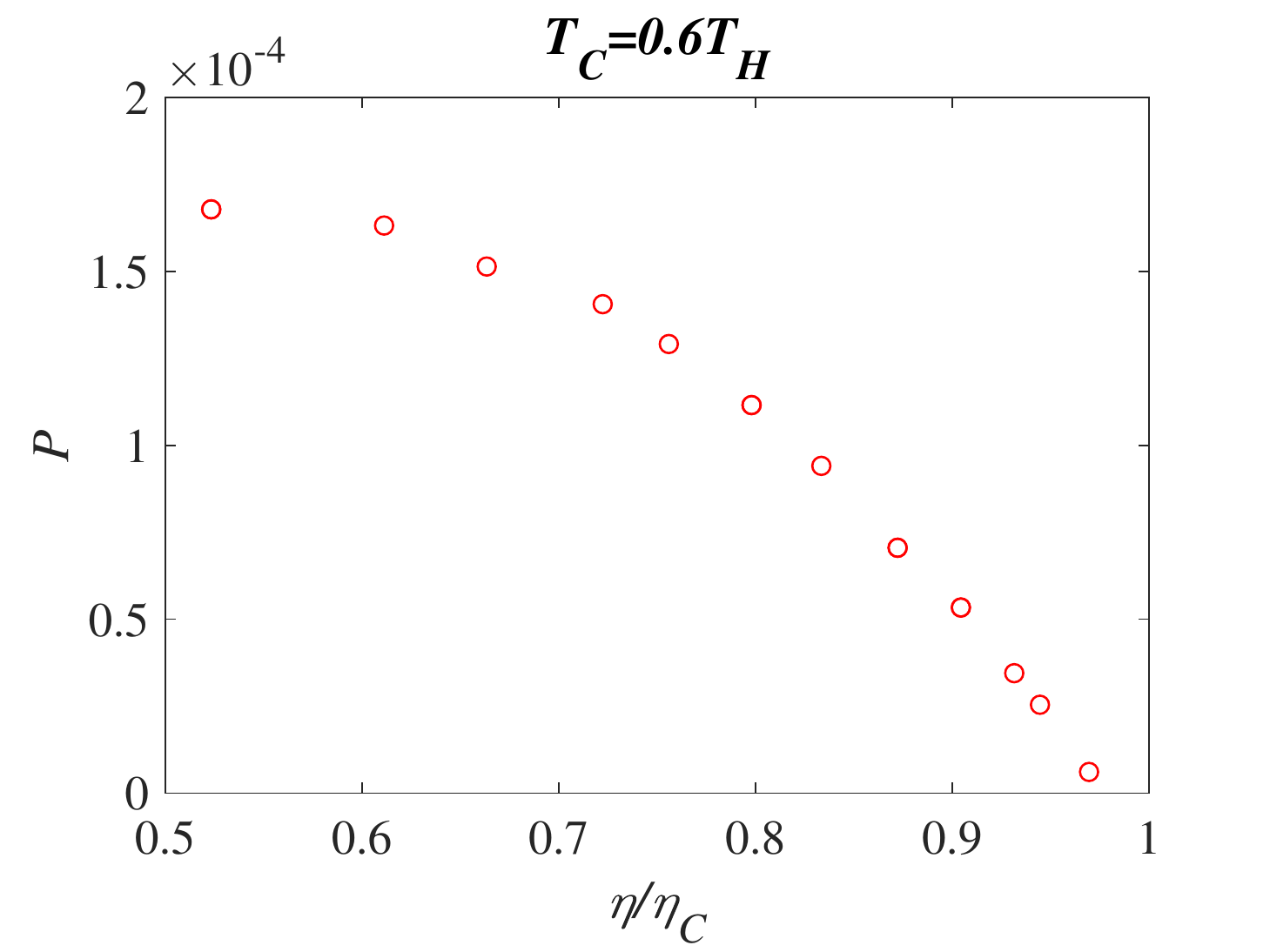}}
	\caption{\label{fig:optimal}
		Pareto front for optimal performance of three-terminal CCQD. 
		(a) $T_C=0.2T_H$, (b) $T_C=0.6T_H$.  }
\end{figure}

So, we briefly scan the whole region to find the proper regime in terms of above three parameters for rapid and unrepeatable optimization process. The result is shown in \Fref{fig:P-V-U}. As you can see, the proper regime of parameters is about $U\in 1.5\sim6T_H$, $\Gamma\in 0\sim0.5T_H$ and $eV\in 0.2\sim0.8T_H$, respectively. Note that, bias voltage is given by unit of $\eta_CU$ and should be converted to the same unit as $U$ and $\Gamma$ in terms of $T_H$ before optimization. Now we can optimize the performance of three-terminal CCQD by using MOGA. The Pareto front and corresponding parameters and objective functions as the result of MOGA optimization are shown in \Fref{fig:optimal} and Table \ref{tab:optimal}. 

In the Pareto front, each point has their own advantage in terms of output power and efficiency. At a glance, one can choose some of them for maximum output power, however, this choice is not always best. Because of relative lower efficiency of these points, more energy costs to maintain the original state. As well-known, the lower the efficiency is, the more energy is delivered to cold reservoir. It can be find easily from Table \ref{tab:optimal}, while the delivered energy from hot to cold reservoir is $U-eV$. So, the more energy is needed to maintain the original temperature difference between hot and cold reservoir. From this point of view, the points with relative higher efficiency have their own advantage, even though they provide us with relative lower output power. 

\begin{table}[t]
	\centering\small
	\begin{tabular}{ccc|ccccccc|cc} 
		\multicolumn{5}{c}{\textsc{$T_C=0.2T_H$}}  &  &
		&\multicolumn{5}{c}{\textsc{$T_C=0.6T_H$}}  \\ 
		\textit{$\Gamma$}  &\textit{U}  &\textit{eV}  &\textit{P}[$10^{-4}$] &\textit{$\eta$} &  &
		&\textit{$\Gamma$}  &\textit{U}  &\textit{eV}  &\textit{P}[$10^{-5}$] &\textit{$\eta$} \\	\hline  
	0.06	&2.3	&1.8	&1.26	&0.778	&	&	&0.02	&2.5	&1	&0.61	&0.388	\\
	0.06	&2.2	&1.7	&2.48	&0.767	&	&	&0.02	&2.6	&1	&3.45	&0.373	\\
	0.06	&2.6	&2	    &2.63	&0.765	&	&	&0.02	&2.4	&0.9	&5.34	&0.362	\\
	0.07	&2	    &1.5	&4.50	&0.744	&	&	&0.03	&2.2	&0.8	&7.05	&0.349	\\
	0.06	&2.3	&1.7	&5.53	&0.733	&	&	&0.03	&2.3	&0.8	&9.41	&0.333	\\
	0.06	&2.2	&1.6	&6.41	&0.721	&	&	&0.03	&2.4	&0.8	&11.16	&0.319	\\
	0.06	&2.1	&1.5	&7.23	&0.707	&	&	&0.03	&2.2	&0.7	&12.91	&0.302	\\
	0.07	&2.3	&1.6	&7.64	&0.689	&	&	&0.03	&2.3	&0.7	&14.06	&0.289	\\
	0.07	&1.9	&1.3	&8.46	&0.676	&	&	&0.03	&2.5	&0.7	&15.14	&0.265	\\
	0.07	&1.9	&1.2	&9.76	&0.622	&	&	&0.04	&2.3	&0.6	&16.32	&0.244	\\
	0.07	&1.9	&1.1	&10.04	&0.568	&	&	&0.04	&2.2	&0.5	&16.78	&0.209	
	
	\end{tabular}
	\centering\caption{Parameters and objective functions  corresponding to the Pareto front at different temperature.}
	\label{tab:optimal}
\end{table}


\section{Conclusions}\label{sec:conclusions}

In this paper we have parametrically investigated the optimal three-terminal quantum heat engine with Coulomb-coupled double quantum dots. We used the master equation and selected the important parameters by varying them and comparing the results. As a matter of fact, these parameters can be controlled in experiment as shown in \cite{N854, Comptes}. Due to their importance and possibility of external control, we have taken GA optimization by using above three chromosome parameters for maximum output power and efficiency. The result is given by the Pareto front and their corresponding parameters and objective functions. Every point in Pareto front has its own advantage and which is best can be chosen in experiments or engineering.  

\ack{
	We wish to thank Dr. OkSong An for kind help. This work is supported by the National Program on Key Science Research of DPR Korea (Grant No. 18-1-3).
}


\appendix
\section{Cotunneling rate and regularization} \label{app:regularization}

Following the lines of \cite{Bruus, 115414}, the cotunneling rates are obtained as 
\begin{eqnarray}
	\eqalign{
	\tilde{\gamma}_{02}^L&=\tilde{\gamma}_{20}^R=\tilde{\gamma}_{gc}^L=\tilde{\gamma}_{cg}^R  \cr
		&=\frac{\Gamma_H}{2\pi\hbar}\int_{-\infty}^{\infty}d \epsilon 
	\left| \frac{1}{\epsilon-\epsilon_d+i\eta}-\frac{1}{\epsilon+\epsilon_d-i\eta} \right| ^2 
	\Gamma_R(\epsilon) f_H(\epsilon) [1-f_R(\epsilon)] \cr
	\tilde{\gamma}_{02}^R&=\tilde{\gamma}_{20}^L=\tilde{\gamma}_{gc}^R=\tilde{\gamma}_{cg}^L  \cr
		&=\frac{\Gamma_H}{2\pi\hbar}\int_{-\infty}^{\infty}d \epsilon 
	\left| \frac{1}{\epsilon-\epsilon_d+i\eta}-\frac{1}{\epsilon+\epsilon_d-i\eta} \right| ^2
	\Gamma_R(\epsilon) [1-f_H(\epsilon)] f_R(\epsilon).
	  \label{eq:cot rate}}
\end{eqnarray}

Inserting the above relation in Eq. \eref{eq:I_LR}, the cotunneling term leads to

\begin{eqnarray}
	\eqalign{ \label{eq:Delta}
		\Delta=\tilde{\gamma_{02}^L}-\tilde{\gamma_{02}^R} \cr
		=\frac{\Gamma_H}{2\pi\hbar}\int_{-\infty}^{\infty}d \epsilon 
		\left| \frac{1}{\epsilon-\epsilon_d+i\eta}-\frac{1}{\epsilon+\epsilon_d-i\eta} \right| ^2
		\Gamma_R(\epsilon) [f_H(\epsilon)- f_R(\epsilon)]. }
\end{eqnarray}

In order to regularize the divergent part, the regularization method by Koch\cite{Koch} has been applied. In this method the divergent integrals are expanded into inverse powers of $\eta$ through the analytic continuation and zero order terms can be represented by polygamma functions using the following relations:
\begin{eqnarray}
\eqalign{
	\sum_{n=1}^\infty \frac{1}{(n+a)(n+b)}=\frac{1}{b-a}\left[\psi(1+b)-\psi(1+a)\right] \cr
	\sum_{n=1}^\infty \frac{1}{(n+a)^2}=\psi^{(1)}(1+a), \label{eq:polygamma}}
\end{eqnarray}
where $\psi$ and $\psi^{(1)}$ are the digamma and trigamma function respectively. To avoid double count for sequential tunneling current, we remove the $O(1/\eta)-$terms, then Eq. \eref{eq:Delta} reads
\begin{eqnarray}
	\eqalign{
	\Delta=\frac{\Gamma_H \Gamma_C}{2\pi\hbar} \frac{1}{\epsilon_d} \frac{\beta_R}{\beta_0} \Re [
	\psi \left( \frac{1}{2}+\frac{\beta_R}{2\pi i} (\mu_R-\epsilon_d) \right)
	-\psi \left( \frac{1}{2}+\frac{\beta_R}{2\pi i} (\mu_R+\epsilon_d) \right)]  \cr
	-\frac{\Gamma_H \Gamma_C}{4\pi^2\hbar} \Im
	[\beta_0 \psi^{(1)}\left(\frac{1}{2}-\frac{\epsilon_d \beta_0}{2\pi i}\right)
	-\frac{\beta_H^2}{\beta_0} \psi^{(1)}\left(\frac{1}{2}-\frac{\epsilon_d \beta_H}{2\pi i}\right) \cr
	+\frac{\beta_R^2}{\beta_0} [ \psi^{(1)}\left(\frac{1}{2}+\frac{\beta_R}{2\pi i}(\mu_R-\epsilon_d)\right)+
	\psi^{(1)}\left(\frac{1}{2}-\frac{\beta_R}{2\pi i}(\mu_R+\epsilon_d)\right) ] ] .  }
\end{eqnarray}


\end{document}